\documentstyle[epsf,aaspp4]{article}
\def\etal{{\em et al.}}
\def\apj{Ap. J.}
\def\apjl{Ap. J. Lett.}
\def\apjs{Ap. J. Supp.}

\begin{document}

\renewcommand{\thefootnote}{\fnsymbol{footnote}}
\title{On the Detectability of Turbulence and Bulk Flows in
X-ray Clusters\footnote[2]{This paper was written in 2000 prior to
the launch of ASTRO-E, but was never submitted or published due to the
unfortunate loss of that spacecraft.  We release it now without major
changes in anticipation of the launch of ASTRO-E2, which should have
sufficient resolution to see the effects predicted here.}}

\renewcommand{\thefootnote}{\arabic{footnote}}
\author{Rashid A. Sunyaev\altaffilmark{1,2}, 
Michael L. Norman\altaffilmark{1,3},
and Greg L. Bryan\altaffilmark{4}}
\received{}
\accepted{}

\altaffiltext{1}{ Max-Planck-Institut f\"ur Astrophysik,
          Karl-Schwarzschild-Strasse 1, 85748 Garching, Germany;
sunyaev@MPA-Garching.mpg.de}
\altaffiltext{2}{Space Research Institute, Russian Academy of 
Sciences, Profsoyuznaya 84/32, 117810 Moscow, Russia}
\altaffiltext{3}{Physics Department and CASS,
University of California at San Diego, CA 92093, USA; 
mlnorman@ucsd.edu}
\altaffiltext{4}{Department of Physics, University of Oxford, 
Keble Road, Oxford Ox1 3RH; gbryan@astro.ox.ac.uk}

\begin{abstract}
Cooling flows, cluster mergers and motions of galaxies through the
cluster gas with supersonic and sonic velocities must lead to large
scale motions of the intracluster medium (ICM). A high--resolution
numerical simulation of X-ray cluster formation by Norman and Bryan
 predicts cluster--wide turbulence with $v_{turb} \sim 300-600$
~km/s and eddy scales $l_{outer} \sim 100-500$ ~kpc, the larger
numbers being characteristic of turbulence near the virial radius,
while the smaller numbers pertain to the core.
The simulation also predicts the existence of ordered bulk flows in
the core with $v \sim 400$ ~km/s on scales of several hundred kpc.
In this paper we consider the observability of such fluid motions
via the distortions it induces in the CMB via the kinematic SZ effect, 
as well as via Doppler broadening and shifting of metal lines in the 
X-ray spectrum.
We estimate $|\Delta T/T|_{kinematic} <$ few $\times 10^{-6}$ ---
at or below current limits of detectability. However, we find that an
energy resolution of a few eV is sufficient to detect several Doppler
shifted components in the 6.7 keV Fe line in the core of the cluster. 
\end{abstract}
\setcounter{page}{1}

\section{Introduction}

Independent lines of observational evidence show that a large fraction of
nearby clusters of galaxies are dynamically young. These include detection
of substructure in optical and X-ray surveys 
(Geller \& Beers 1982; Dressler \& Shectman 1988, Jones \& Forman 1992), 
non-Maxwellian galaxy velocity distributions in clusters containing
substructure (Beers, Flynn \& Gephardt 1990;
Pinkney \etal ~1996), X-ray surface brightness distortions (Mohr, Fabricant 
\& Geller 1993; Buote \& Tsai 1995), and X-ray temperature substructure
(Henry \& Briel 1992; Markevitch \etal ~1994).
These features can be understood as the result of cluster mergers
(e.g., Roettiger, Burns \& Loken 1996), which are predicted in
hierarchical structure formation scenarios (e.g., White, Briel \& Henry
1993). 

Cluster mergers induce large--scale bulk flows with velocities
of order the virial velocity ($\sim 1000$ ~km/s for rich clusters.) 
Cluster-wide turbulence will then be established in the intracluster 
medium (ICM) in a few turnover times of the largest eddies
$\tau \sim r_{vir}/\sigma_{vir} \sim 10^9 ~yr$. 
We expect the turbulence to possess a Kolmogorov-like spectrum down
to the dissipation scale $l_{diss} \sim l_{outer}/Re^{3/4}$, where
$Re$ is the Reynolds number. Due to the low densities and high
temperatures of the ICM plasma, the Coulomb mean free path is of order 
10-100 kpc,
yielding the {\em classical} estimate $Re \sim 100$. 
However, the presence of a weak magnetic field will reduce transport
coefficients from the Spitzer values by a factor of $\sim 1/M^2$, where
$M \equiv \frac{v}{\sqrt{B^2/4\pi \rho}}$
is the Alfv\'{e}n Mach number (Tao 1995).
Using typical values for X-ray cluster cores(halos) of $v=300(600)$ ~km/s,
$n=10^{-3}(10^{-5})$ ~cm$^{-3}$, and $B=10^{-6}(10^{-8}$) G,
we find $M^2 \sim 10(10^4)$. This increases the effective Reynolds
number to $\sim 10^3(10^6)$, predicting dissipation scales of
$\sim$ 5 kpc(30 pc), assuming $l_{outer} = 1$ Mpc. 

There are two ways to study the nature and importance of these motions
(i.e., their contribution to the local pressure of the gas, mixing of the 
high Z elements in the cluster and their contribution to ion heating due
to dissipation in the smallest scales). First, through numerical simulation 
of the evolution of the gas in the cluster,
taking into account evolution of the gravitational potential,
mergings, heating and cooling of the gas, formation of the cooling
flows etc. Recently Norman \& Bryan (1999) published results of such 
simulations.
Second, to use X-ray spectroscopy missions (under construction now and
under consideration) to measure the real distribution of velocities in
rich clusters of galaxies. We are excited by the progress in the
energy resolution of AXAF and XMM (with X-ray grating), ASTRO-E (with
X-ray bolometers).
CONSTELLATION
and XEUS projects planned for the launch by NASA and ESA in the middle
of the next decade would have 1-2 eV energy resolution in the whole
band from 1 to 7 keV. Such energy resolution will permit measurements of
velocities as low as 50-100 ~km/sec, more than an order of magnitude lower
than the sound speed in rich clusters of galaxies.

In this paper we use the published data on the velocity distribution
in a simulated rich cluster of galaxies (Norman \& Bryan 1999)
to demonstrate that the observations of X-ray lines with high
energy resolution would open the new method to investigate the large scale
intergalactic gas velocity distribution in clusters of galaxies.
It is important for us that we are dealing with heavy
elements and especially with iron, which is 56 times heavier than
hydrogen and therefore its thermal velocity (and corresponding
thermal line broadening) is 7.5 times lower than the 
proton thermal velocity.
This opens the way to measure subsonic turbulent velocities.

Observations and simulations have shown that the gas in the cluster is not 
isothermal (Markevitch \etal ~1998, Frenk et al. 1999). 
Using lines of different ions
and elements we get an opportunity to measure the velocity distribution
in the regions with different temperatures in the same line of sight.
We note that there is only one way to compete with X-ray observations to
measure cluster gas motions. This is to
observe the microwave hyperfine structure lines of heavy ions (for
example 3.03 mm line of Lithium-like iron-57)(Sunyaev \& Churazov,
1994). These lines are analogous to 21 cm line of hydrogen.
Unfortunately they are not sufficiently bright, therefore it is better
to observe the broadening of X-ray lines.

\section{The Numerical Simulation}

We have simulated the formation of an adiabatic X-ray cluster in an
$\Omega = 1$ universe. The initial spectrum of density fluctuations
is CDM-like with a shape parameter of $\Gamma = 0.25$ (Efstathiou
et al. 1992). The cluster itself is a $3\sigma$ fluctuation at the center
of the box
for a Gaussian filter of 10 Mpc radius. We use a Hubble constant 
of 50 km/s/Mpc and a baryon fraction of 10\%. This cluster is the
subject of a comparison project between twelve diffferent simulation
methods, the results of which can be found in Frenk et al. (1999).

The evolution of gas and dark matter is computed using a new method
to numerical cosmology: adaptive mesh refinement (AMR).
AMR utilizes a logical hierarchy of finer resolution
meshes in regions of the calculation which require high resolution
(in our case, collapsing halos of dark matter and gas.) AMR is adaptive,
automatic and recursive: an arbitrary number of sub-meshes of different 
levels of resolution are automatically created as the solution evolves.
Within each sub-mesh, the equations of adiabatic gas dynamics are solved
subject to boundary conditions interpolated from overlying coarser meshes
using the piecewise parabolic method (PPM) as modified for cosmology by 
Bryan et al. (1995). Dark matter dynamics is solved using an adaptive 
particle-mesh (APM) algorithm inspired by Couchman's
(1991) algorithm. Details of the method can be found in Bryan \& Norman
(1997), Norman \& Bryan (1999).

The simulation was initialized with two grids at two levels of
refinement. The first is the root grid covering the 64 Mpc$^3$ triply
periodic domain with $64^3$ cells. The second grid is also $64^3$ cells,
but is only 32 Mpc on a side and is centered on the cluster. Thus,
over the region that forms the cluster, we have an initial cell size
of 500 kpc leading to an approximate mass resolution of $8.7 \times 10^8 
M_{\odot} ~(7.8 \times 10^8 M_{\odot})$ for the baryons ~(dark matter).
Cells are flagged for refinement when the baryon mass exceeds
$4 M_{initial} \approx 3.5 \times 10^9 M_{\odot}$. We use a refinement
factor R = 2, thus immediately after refinement, a refined cell's
baryon mass will be $0.5 M_{initial}$. As the simulation evolves,
as many as $\sim 400$ grids at seven levels of refinement are created,
with a minimum cell size of 64 Mpc/$64 \times 2^7$ = 7.8 kpc.
Our method thus maintains
mass resolution in the gas and dark matter, while providing high
force resolution where needed. In this regard, we accomplish the
same thing as an adaptive smoothing length SPH calculation (e.g.,
Navarro, Frenk \& White 1995) with a less viscous, more accurate 
shock-capturing hydrodynamic scheme. 
This is a distinct advantage in simulations of cluster turbulence.

Figure 1 shows radial profiles of the spherically averaged gas and
dark matter velocity dispersions in the cluster at z = 0. The cluster
centroid is taken to be the point of maximum total density.  
The dark matter velocity dispersion
is computed after subtracting off the mean cluster peculiar velocity
inside a sphere of radius $r_{200} \approx 2.2 Mpc$. It shows the
characteristic quasi-isothermal plateau $\sigma \approx \sigma_{vir}$
for $0.05 \leq r/r_{vir} \leq 0.5$, sharply declining in the outer 
and inner parts of the cluster as discussed by Frenk et al. (1998). 
The gas velocity dispersion exhibits a different profile as the
dark matter as the kinetic temperature contribution is not included.
Only peculiar fluid velocities due to bulk motions or turbulence
are reflected here. We compute the fluid velocity dispersions
relative to the mean {\em fluid} peculiar velocity inside spheres
of three different radii: 0.01, 0.1, and 1 $r_{vir}$. The
curves for 0.1 and 1 $r_{vir}$ are nearly identical, and show
fluid peculiar velocities are dominated by infall for $r \geq r_{vir}$.,
Inside the main cluster shock at $r \approx r_{vir}$, the ICM is in a 
turbulent state (Bryan \& Norman 1998, Norman \& Bryan 1999), 
with $\sigma_{gas}$ declining from 800 km/s to $\sim$ 400 ~km/s at
$r = 0.5 r_{vir}$. The turbulent velocities are roughly constant 
inside $0.5 r_{vir}$, which is the significant new result of our
simulation. Visual inspection of the flowfield reveals turbulent
eddies with a range of sizes 50 - 500 kpc. In addition, the flowfield
in the central few hundred kpc exhibits an ordered circulation with
$v \approx 400$  km/s. Visualizations of the velocity 
field can be found in Norman \& Bryan (1999) and at the web site
http://zeus.ncsa.uiuc.edu:8080/Xray/clusters.html.

In the following sections, we explore whether such fluid motions are
detectable.

\begin{figure}
\epsfysize=6in
\centerline{\epsfbox{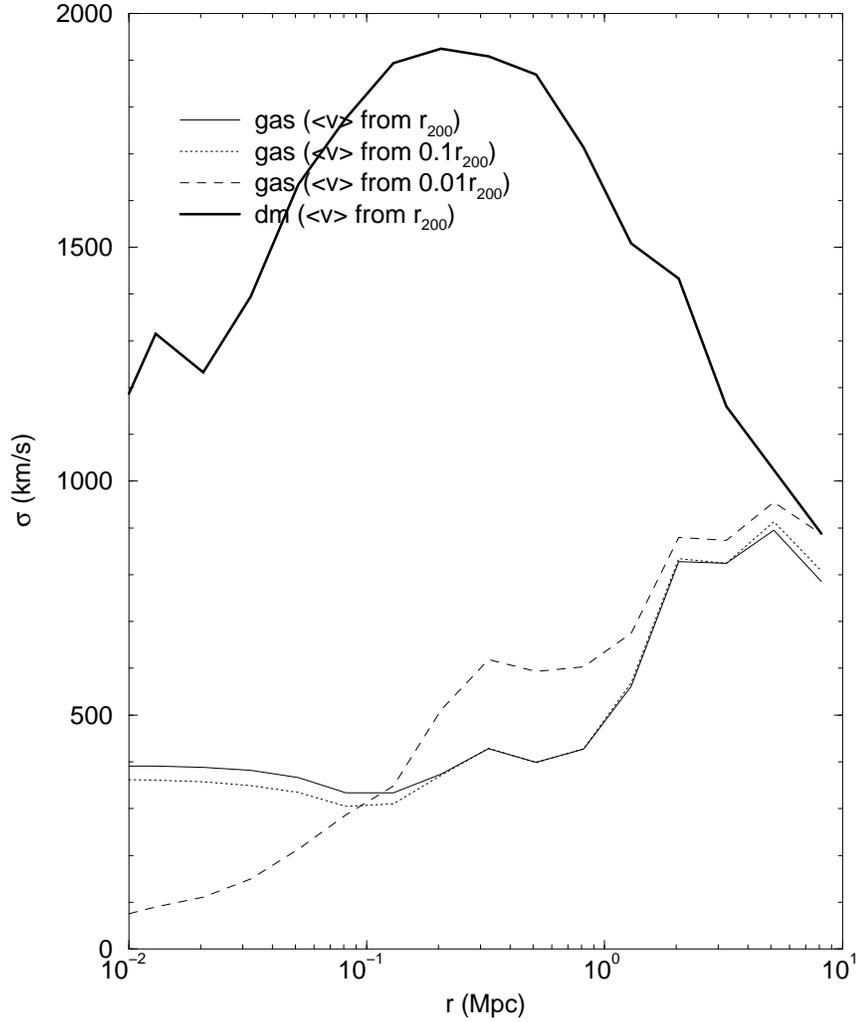}}
\caption{
Spherically averaged profiles of gas (thin lines) and dark matter
(heavy line) 3D velocity dispersions in the simulated 
X-ray cluster.
The gas velocity dispersions are computed relative to the center-of-mass
velocities inside spheres of different radii: ($r_{200}$ (thin solid line);
$0.1 r_{200}$ (thin dotted line); and $0.01 r_{200}$ (thin dashed line).
The plateau in $\sigma_{gas} \sim 400 ~km/s$ for $r \leq 1 Mpc$ is
indicative of bulk velocities/turbulence in the cluster gas. 
}
\label{fig:gas_rms}
\end{figure}

\section{Results}

\subsection{Kinematic and Thermal SZ Effect}

A hot ICM will inverse Compton scatter the cosmic microwave background
radiation, boosting photons to higher frequencies and distorting the
Planck blackbody spectrum (Sunyaev \& Zel'dovich 1970, 1980). In the
Rayleigh-Jeans part of the spectrum, the temperature decrement due to
this effect is $\Delta T/T = -2y$, where
$y$ is the line of sight integral of essentially the gas pressure 
through the ICM

\begin{equation}
y = \int_{-\infty}^{\infty} \frac{kT_e}{m_e c^2} \sigma_T n_e d\ell.
\end{equation}
\noindent
$T_e, m_e,$ and $n_e$ are the electron temperature, mass, and number
density and $\sigma_T$ is the Thompson scattering cross section.
In addition to this {\em thermal} effect, which is insensitive
to fluid motions in the ICM, there is also a {\em kinematic} effect,
which to first order in $\frac{v}{c}$ is 
\begin{equation}
\frac{\Delta T}{T} = -\frac{1}{c} \int_{-\infty}^{\infty} \sigma_T n_e 
v_{\parallel} d\ell,
\end{equation}
\noindent
independent of frequency. 
Here $v_{\parallel}$ is the line-of-sight component of the cluster 
gas peculiar velocity (i.e., relative to the CMB frame.)
The total temperature decrement is the sum of these two effects
\begin{equation}
\left.\frac{\Delta T}{T}\right)_{tot} = 
\left.\frac{\Delta T}{T}\right)_{thermal} + 
\left.\frac{\Delta T}{T}\right)_{kinematic} = 
-\tau_T \left[2\frac{k<T_e>}{m_e c^2} + \frac{<v_{\parallel}>}{c}\right] ,
\end{equation}
\noindent
where $\tau_T$ is the Thompson optical depth, and $<T_e>$ and $<v_{\parallel}>$ are the electron density-weighted electron temperature and fluid velocity
along the line-of-sight. 

Because of the $n_e$ weighting and the fact that
observed X-ray clusters have monotonically decreasing
temperature profiles (Markevitch et al. 1998), we expect the
thermal SZ effect to be highest in the core. 
Figures 2a and b confirm this expectation, where we present images of the
y-parameter for our simulated cluster on two spatial scales. 
Isocontours of $y$ are
smooth, concentric, and nearly circular, consistent with recent
radio interferometric observations (Carlstrom et al. 1999). The peak
value of $y \approx 10^{-4}$ is also consistent with observed values 
(Carlstrom et al. 1999), although our result is naturally sensitive 
to our assumed $\Omega_b$.

Since $v_{\parallel}$ is a signed quantity, the magnitude of
the kinematic SZ effect will depend not only on the strength of
turbulent or bulk motions in the cluster core, but also on the number of 
velocity reversals along the line of sight. Images of the kinematic
SZ effect on two spatial scales are shown in Figs. 2c and d, obtained by 
integrating equation (2) along multiple lines-of-sight through the simulated 
cluster. One sees a reversal in $\Delta T /T$, due to the aforementioned
circulation in the cluster core. Across the center of the cluster,
$\Delta T /T$ changes abruptly from -4 to +4 $\times 10^{-6}$, or about 
2\% the magnitude of the thermal effect. Let us check this result for
consistency. From equation (3), setting $k<T_e>$ = 4.8 keV (the mass weighted
temperature inside $r_{vir}$), we deduce $<v_{\parallel}> = 56$ ~km/s,
about a quarter of the one-dimensional gas velocity dispersion in the
core 
In addition to the dominant effect due to core circulation, we can 
see 
cluster--wide fluctuations in the kinematic SZ effect image due
to turbulent motions in the ICM, as well as due to a subcluster
falling in at 3 O'clock in Fig. 3c. These fluctuations are at an
amplitude of $\Delta T /T \approx 10^{-6}$. 

The kinematic SZ effect due to cluster turbulence is just below current
limits of detectability. Subtraction of background radio galaxies,
which are not resolved in the BIMA and OVRO interferometers
used for the current observations, introduce a systematic uncertainty of 
15-25 $\mu$K (Cooray et al. 1999) in the SZ decrement, 
versus our peak signal of $\sim 11 \mu$K. Better point source subtraction
using future arrays with higher spatial resolution and visibility
coverage should allow us to detect the kinematic effect due to
cluster mergers, turbulence and bulk flows in the ICM. 

\begin{figure}
\epsfysize=5in
\centerline{\epsfbox{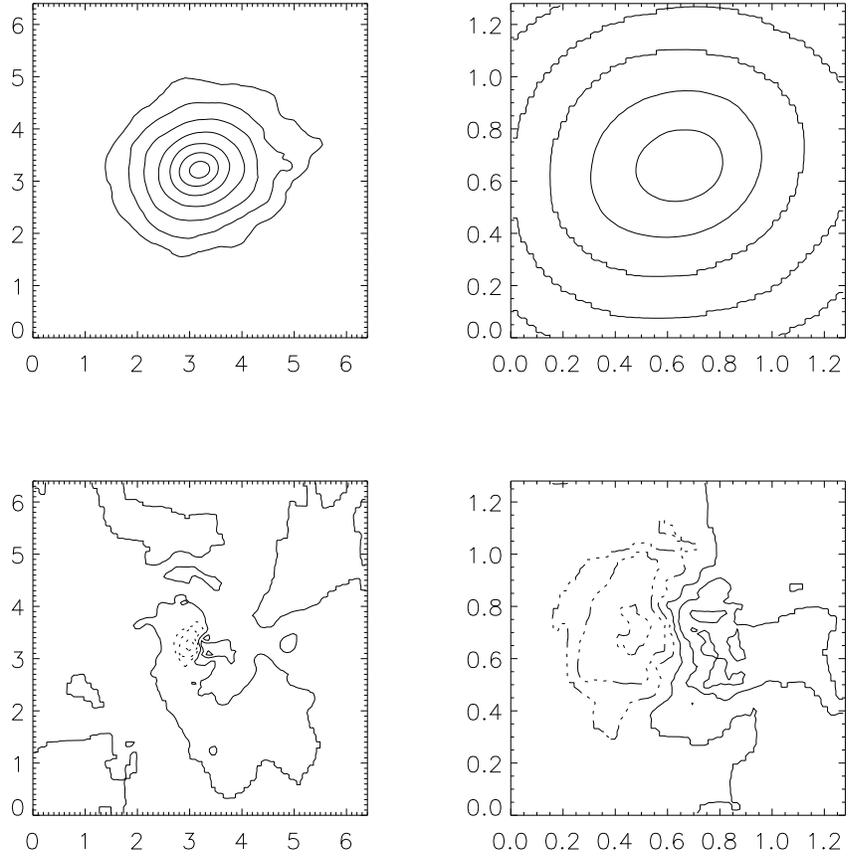}}
\caption{
Top: the SZ thermal y-parameter for a region
6.4 Mpc (left) and 1.2 Mpc (right) on a side.  Bottom: the
SZ kinematic effect, in terms of $\Delta T/T$.
The y contours are logarithmically spaced:
1, 2, 4, 8, 16, 32 and 64 x $10^{-6}$ for the large (6.4 Mpc) region
and   4, 8, 16, 32 and 64 x $10^{-6}$ for the small (1.2 Mpc) region.
The kinematic contours are -4, -2, 1, 0, 1, 2, 4 x $10^{-6}$ and
the negative contours are dotted.
}
\label{fig:SZ_contour}
\end{figure}

\begin{figure}
\epsfysize=5in
\centerline{\epsfbox{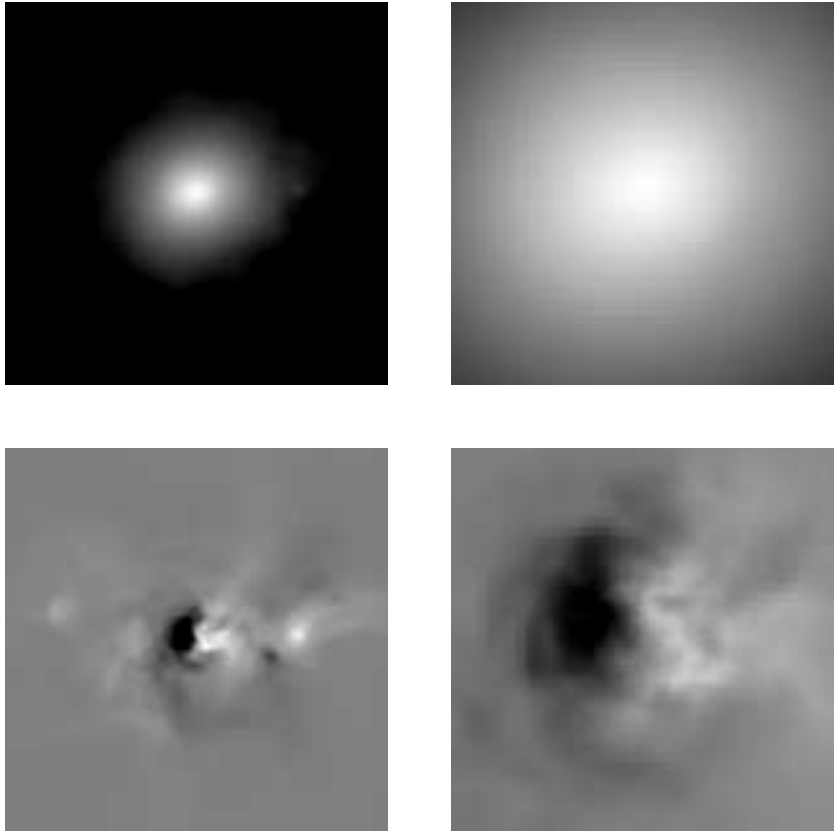}}
\caption{
Same as previous figure, but an image representation.
The y-parameter is log-scaled from $10^{-6}$ to $10^{-4}$ (both images),
and the kinematic $\Delta T/T$ goes from -2 to +2 x $10^{-6}$ on for the large
(left) image and -4 to +4 x $10^{-6}$ for the small image (1.2 Mpc, on the
right).
}
\label{fig:SZ_image}
\end{figure}

\subsection{Doppler Broadening and Shifting of Metal Emission Lines}

The radial component of the thermal velocity of iron ions is $\sim 130$
km/s for 5 keV temperature. Radial turbulent velocities decrease from 
$\sim 500$ ~km/s at the virial radius to $\sim 250$ km/s in the core 
(Fig. 1). In addition, we have radial infall of gas and subclusters
beyond the virial radius with $|v_r| > 1000 km/s$, and an ordered circulation
with $v \sim 400$ km/s in the core. Thus we might expect to see in a
high resolution X-ray spectrum multiple components in iron emission
lines with Doppler shifts of magnitude $\Delta E$ 2-6 times the thermal
width, or $\pm 6-18$ eV. We examine this possibility quantitatively in Fig. 4.

To isolate the effects of gas motions on X-ray line spectra,
we plot the distribution of emission for an infinitely thin
line with intensity proportional to $n_e^2f(T_e)$ versus radial velocity
relative to the cluster rest frame (top axis) or energy (bottom axis)
for nine lines-of-sight through the simulated cluster.
The line emissivity is assumed to be proportional to $n_e^2f(T_e)$.
In Fig. 4 we set $f(T_e)=1$ for simplicity; 
below we construct more realistic spectra.
Each l.o.s. is normalized to the total amount of flux in
the line, and that normalization (relative to the line through the
cluster center) is shown in the upper right-hand corner.  

In the absence of bulk motions and turbulence, the plot should show 
delta function emission at $\Delta E = 0.$ Instead we see multiple
components spread over the range of energies estimated above. The
asymmetric profiles seen at negative impact parameters are the result
of a large sub-clump falling into the cluster at high speed. The
profiles at impact parameters $ b \geq 0$ show a roughly symmetric
plateau of components over the range $\pm 15$ eV, with the exception
of a strong peak at $b=0, \Delta E = -5$ eV caused by fluid circulation
in the core.
These componets  will be observable with order 1 eV spectral resolution, 
although point out that each
l.o.s. is pencil-thin and that a real observation will be the
(flux-weighted) average of many of these.  The net result I imagine
will be a much broader line with large sub-clumps moving at high speed
showing up more clearly.  

\begin{figure}
\epsfysize=6in
\centerline{\epsfbox{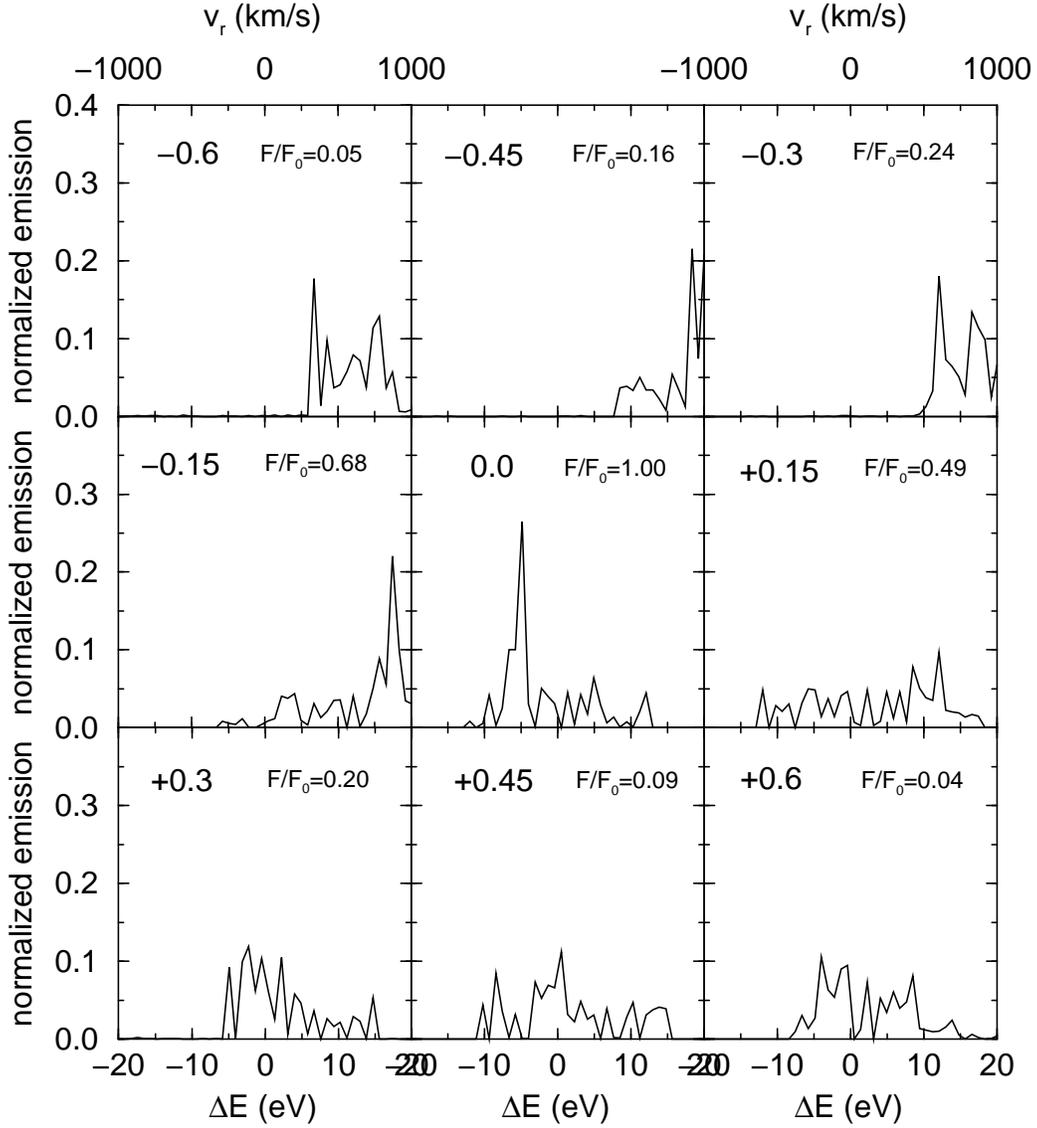}}
\caption{
The normalized X-ray emission versus Doppler shift in eV
along nine lines of sight through the simulated X-ray cluster. 
The corresponding radial velocities are given at the top of the plot.
The impact parameter in Mpc for each l.o.s is given at upper left in each panel.
The flux $F$ in units of the central flux $F_o$ is given
at upper right in each panel.
}
\label{fig:splot_nine}
\end{figure}

\begin{figure}
\epsfysize=6in
\centerline{\epsfbox{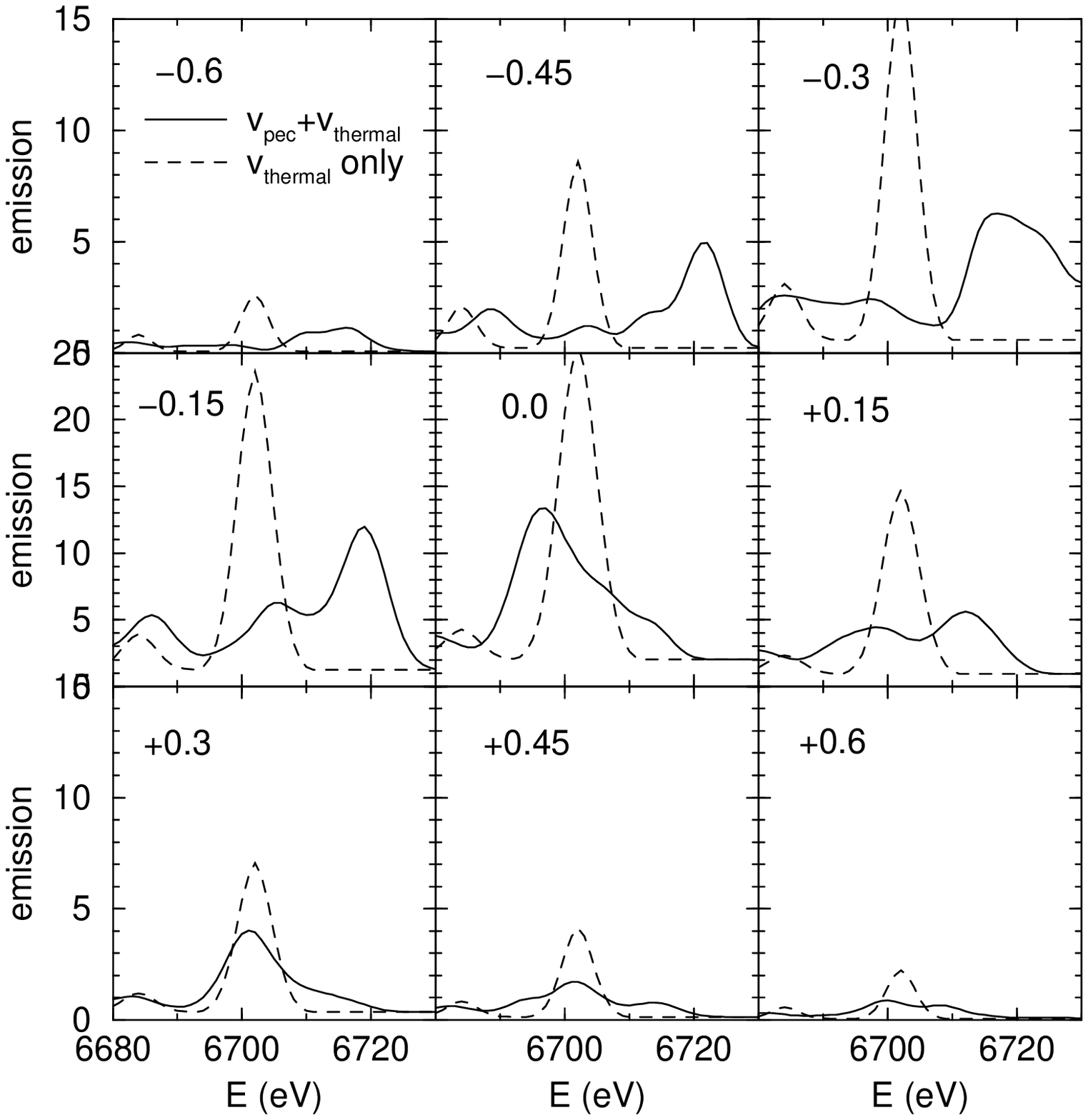}}
\caption{
Synthetic Fe line spectra along nine lines of sight through the
simulated X-ray cluster. The dashed lines are computed assuming
thermal broadening only, whereas the solid lines include both
thermal broadening and Doppler shift.
}
\label{fig:SZ_image}
\end{figure}

We now investigate what a realistic spectrum would look like, including
the effects of thermal broadening, Doppler broadening and shifting
due to fluid motions, and temperature effects along the l.o.s.
Fig. 5 shows the 6.702 keV iron line as predicted
by a Raymond-Smith code. We assume a constant iron abundance.
The lines-of-sight are identical to Fig. 4. 
The dashed line shows the effect of thermal broadening only, while the
solid line shows the full effect of thermal plus bulk motions.
The emission is not normalized in this plot, as can be seen from
the different scales at left. The absolute scale is arbitrary
as we have not defined the cross-section of the line of sight.

Again, the infalling sub-cluster is visible as a shifted maximum, 
particularly in the  b = -0.3 plot. 
At b = 0 and +0.15, the spectrum is turbulently broadened to several
times its thermal width. Discrete subcomponents are still visible.
Finally, the second peak seen at 6.685 keV is actually
another iron line (in fact there are a whole series at lower energy,
but this is clearly the strongest).  The energy resolution in this
plot is much higher than any planned mission, but it is evident
that ~4 eV resolution (such as that planned for ASTRO-E2) would be 
enough to see these nonthermal features, 

\section{Discussion and Conclusions} 

The preceeding results suggest that X-ray spectroscopy is the most
promising approach in the near term to detect bulk motions and
turbulence in X-ray clusters. Therefore we focus our discussion on these
possibilities. 


Our hydrodynamical calculations show that the dynamics of cluster formation in
a CDM-dominated model produce cluster-wide turbulence and ordered
bulk flows in the ICM which are strong enough to be detected
in the cluster core in high resolution X-ray spectra.
Due to the small thermal broadening of emission lines from heavy elements, 
these are the most promising probes.
We have calculated a synthetic spectrum of the well-separated
6.702 keV emission line of iron which shows strong turbulent broadening
and shifting of the line into multiple components over a range $\Delta E \sim
\pm 15$ eV. How reliable is this result?
Our finite numerical resolution in the core ($\sim 15$ kpc)
artificially cuts off the turbulent spectrum at this scale. Thus
our simulation underestimates turbulent broadening due to microturbulence  
which could possibily eliminate
discrete, Doppler shifted components. However, we have seen that some
components are caused by ordered motions in the core, as well as 
infalling sub-clusters, and our resolution is adequate to predict
these features with some confidence. 
A second effect not included in our synthetic spectrum is
finite angular resolution, which would average over components making
spectra more featureless. 
Thus, we expect X-ray spectra with a resolution of
a few eV will see metal lines whose widths are dominated by turbulent
broadening, with possibly a few components reflecting bulk motions in the
core. This implies that temperature measurements from line widths will be 
difficult or impossible, as they will be dominated by turbulent broadening.
Fig. 5 demonstrates this is true, not only in the center, but out
to several core radii.  However, since the turbulent velocity scales 
with the virial velocity of the cluster, line widths will probe the 
depth of the potential well. 


Our predictions are based on a numerical simulation which makes
a number of simplifying physical assumptions about the thermal and dynamic
properties of the gas which may not reflect conditions in real X-ray
clusters. For example, we ignore radiative cooling, which is important
in the majority of real X-ray clusters at low redshift. In the simplest
models radiative cooling in dense cluster cores
lead to centrally directed cooling flows and lower
central temperatures. There is some observational support for these
models (Fabian 1994). However, the velocity and temperature structure
in the center of a cooling flow cluster is likely to be considerably
more complicated, especially if the medium is turbulent or recovering
from a recent merger. It is clear
from our results that high angular and spectral resolution X-ray 
spectroscopy will be a powerful tool to probe the physical nature
of cooling flows clusters. We can imagine that turbulence provides
the perturbations which, when amplified by thermal instability,
yield a two-phase medium in cooling flows. 
Lines of S and Ar would be brighter in the cooler gas. 
Mapping the cluster in these lines would be very revealing.  
We also assume the electron and ion temperatures are in LTE.
Recently, Chieze \etal ~(1998) have shown via 3D simulations
that this is a poor assumption near the virial radius. Since
the line emissivity is sensitive to the electron temperature,
while the line widths in the absence of turbulent broadening
are a functionof the ion temperature, one could {\em in principle}
determine both with sufficiently accurate observations. However,
the low emission measure and high level of turbulence we find
at large radii would make this measurement difficult if not
impossible.





\noindent
{\em Acknowledgements} MLN would like to thank the gracious hospitality
of Simon White at the Max-Planck-Institut f\"{u}r Astrophysik where this
work was done, and the Alexander von Humboldt Foundation for financial
support during my stay. The numerical simulations were carried out
on the SGI/CRAY Origin2000 system at the National Center for Supercompting
Applications, University of Illinois at Urbana-Champaign. This
project is partially supported by NSF grant AST-9803137 and NASA grant
NAGW-3152.



\begin{thebibliography}{}

\bibitem{}
Beers, T. C., Flynn, K., Gebhardt, K. 1990, AJ, 100, 32

\bibitem{}
Bryan, G.L. and Norman, M.L. 1997. in {\em Computational Astrophysics}; 
12th Kingston Meeting on Theoretical Astrophysics, eds. D. Clarke \& M. West, 
ASP Conference Series Vol. 123, 363

\bibitem{} 
Bryan, G.L. and Norman, M.L.. 1998. \apj, 495, 80

\bibitem{}
Bryan, G.L., Norman, M.L., Stone, J.M., Cen, R. and Ostriker, J.P. 1995.
{\em Comput. Phys. Comm.}, {\bf 89}, 149

\bibitem{}
Buote, D.A. \& Tsai, J.C. 1995. \apj, 452, 522

\bibitem{}
Carlstrom, J. E.; Mohr, J. J.; Reese, E. D.; Holder, G. P.;
 Leitch, E. M.; Joy, M. K.; Grego, L.; Patel, S.; Holzapfel, W. L.
1999. BAAS, 194, 5801

\bibitem{}
Chieze, J.-P.; Alimi, J.-M.; Teyssier, R. 1998. \apj, 495, 630

\bibitem{}
Cooray, A. R.; Carlstrom, J.E.; Grego, L.;
 Holder, G. P.; Holzapfel, W. L.; Joy, M.;
 Patel, S. K.; Reese, E. 1999. 
in ``After the Dark Ages: When Galaxies were Young (the Universe at 2 < z < 5)"
eds. S. Holt \& E. Smith. AIP Press, Washington DC, 1999, p. 184

\bibitem{}
Couchman, H. M. P. 1991. \apjl, 368, 23

\bibitem{}
Dressler, A. \& Schectman, S.A. 1988. AJ, 95, 985

\bibitem{}
Efstathiou, G., Bond, J. R. \& White, S. D. M. 1992. MNRAS, 258, 1

\bibitem{}
Fabian, A. C. 1994. ARA\&A, 32, 277

\item Frenk, C.S., White, S.D.M., Bode, P., Bond, J.R., Bryan, G.L., Cen,
R., Couchman, H.M.P., Evrard, A.E., Gnedin, N., Jenkins, A., Khokhlov,
A.M., Klypin, A., Navarro, J.F., Norman, M.L., Ostriker, J.P., Owen, J.M.,
Pearce, F.R., Pen, U.-L., Steinmetz, M., Thomas, P.A., Villumsen, J.V.,
Wadsley, J.W., Warren, M.S., Xu, G. and Yepes, G. 1999. \apj, 525, 554

\bibitem{}
Geller, M. J.; Beers, T. C. 1982. PASP, 94, 421

\bibitem{}
Henry, J. P.; Briel, U. G. 1993. Advances in Space Research, vol. 13, no. 12, 
p. (12)191-(12)198 

\bibitem{}
Jones, C. \& Forman, W. 1992. in ``Clusters and Superclusters of Galaxies", 
NATO Advanced Science Institutes (ASI) Series C, Volume 366, Ed. A. C. Fabian, 
Dordrecht: Kluwer, 49

\bibitem{}
Loken, C., Norman, M.L., Nelson, E., Burns, J.O., Bryan, G., \& 
Motl, P. 2002. \apj, 579, 571

\bibitem{}
Markevitch, M., Forman, W.R., Sarazin, C.L. \& Vikhlinin, A. 1998.

\bibitem{}
Markevitch, M., Yamashita, K., Furuzawa, A. \& Tawara, Y. 1994. \apjl, 436, 71
\apj, 503, 77

\bibitem{}
Mohr, J.J., Fabricant, D.G. \& Geller, M.J. 1993. \apj, 413, 492

\bibitem{}
Navarro, J.F., Frenk, C.S. \& White, S. D. M. 1995. MNRAS, 275, 720

\bibitem{}
Norman, M.L. \& Bryan, G.L., 1999. 
in {\em Numerical Astrophysics 1998}, eds. S. Miyama \& K. Tomisaka, 
Astrophysics \& Space Science Library Vol. 240, (Kluwer, Boston), 19

\bibitem{}
Norman, M.L. \& Bryan, G.L. 1999. in 
{\em The Radio Galaxy Messier 87}, eds. H.-J. Roeser \& K. 
Meisenheimer, Lecture Notes in Physics No. 530, (Springer,
Heidelberg), 106

\bibitem{}
Pinkney, J., Roettiger, K., Burns, J. O. \& Bird, C. M. 1996. \apjs, 104, 1

\bibitem{}
Roettiger, K., Burns, J.O. \& Loken, C. 1996. \apj, 473, 651

\bibitem{}
Sunyaev, R. \& Churazov, E. 1994. MNRAS, 297, 1279

\bibitem{}
Sunyaev, R. A. \& Zeldovich, Ya. B. 1970. Ap\&SS, 7, 3

\bibitem{}
Sunyaev, R. A. \& Zeldovich, Ya. B. 1980. ARA\&A, 18, 537

\bibitem{}
Tao, L. 1995. MNRAS, 275, 965

\bibitem{}
White, S. D. M., Briel, U. G. \& Henry, J. P. 1993. MNRAS, 261, L8

\end{thebibliography}
\end{document}